# Pumping laser excited spins through MgO barriers

Ulrike Martens[1], Jakob Walowski[1], Thomas Schumann[1], Maria Mansurova[1], Alexander Boehnke[2], Torsten Huebner[2], Günter Reiss[2], Andy Thomas[3] and Markus Münzenberg[1]

1 Institut für Physik, Ernst-Moritz-Arndt Universität Greifswald, Felix-Hausdorff-Straße 6, 17489 Greifswald, Germany

2 Centre for Spinelectronic Materials and Devices, Physics Department, Bielefeld University, Universitätsstraße 25, 33615 Bielefeld, Germany

3 Leibniz Institute for Solid State and Materials Research Dresden (IFW Dresden), Institute for Metallic Materials, Helmholtzstrasse 20, 01069 Dresden, Germany

E-Mail: ulrike.martens@uni-greifswald.de

**Abstract.** We present a study of the tunnel magneto-Seebeck (TMS)[1] effect in MgO based magnetic tunnel junctions (MTJs). The electrodes consist of CoFeB with in-plane magnetic anisotropy. The temperature gradients which generate a voltage across the MTJs layer stack are created using laser heating. Using this method, the temperature can be controlled on the micrometer length scale: here, we investigate, how both, the TMS voltage and the TMS effect, depend on the size, position and intensity of the applied laser spot. For this study, a large variety of different temperature distributions was created across the junction. We recorded two-dimensional maps of voltages generated by heating in dependence of the laser spot position and the corresponding calculated TMS values. The voltages change in value and sign, from large positive values when heating the MTJ directly in the centre to small values when heating the junction on the edges and even small negative values when heating the sample away from the junction. Those zero crossings lead to very high calculated TMS ratios. Our systematic analysis shows, that the distribution of the temperature gradient is essential, to achieve high voltage signals and reasonable resulting TMS ratios. Furthermore, artefacts on the edges produce misleading results, but also open up further possibilities of more complex heating scenarios for spincaloritronics in spintronic devices.

## 1. Introduction

The demand for new concepts to advance progress in information processing calls for technologies that not solely rely on the electrons charge, but also control the electrons other property, the spin. In the research field of magnonics those concepts are approached by making use of spin-waves [1]. This method opens new aspects for information processing [2, 3]. Gaining control over the excitation and propagation of excited spins is in the focus, both for potential magnetic storage and logic devices. The investigation of spincaloritronic effects is one of the key elements to understand the underlying principles and develop the required techniques. The TMS is defined as the change of the Seebeck coefficients ($S_\mathrm{p}, S_\mathrm{ap}$) in MTJs regarding the parallel (p) and antiparallel (ap) magnetic configuration of the ferromagnetic electrodes [4–6] while creating a temperature gradient across the layer stack. The resulting temperature difference between the two electrodes leads to the generation of a Seebeck voltage. Because the driving force to generate voltages are temperature gradients, this method also provides the opportunity to recycle waste heat usually generated in electronic spintronic devices. In recent spincaloritronic research, several groups have observed the TMS effect in MTJs with CoFe electrodes and MgO tunnel barriers [4, 5]. CoFeB/MgO/CoFeB tunnel junctions are well established devices. In 2001, it was predicted that crystalline MgO as barrier material provides high tunnel magnetoresistance (TMR) ratios due to the coherent tunneling of fully spin-polarized $\Delta_1$ electrons [7, 8]. The largest reported TMR ratio measured experimentally at room temperature was by Ikeda et al. with a value of 604 % [9]. Furthermore, this class of tunnel junctions is relatively easy to prepare. The material layers can be deposited using sputtering techniques. For patterning optical mask lithography is sufficient. Consequently, the preparation times are rather short. Accordingly, the investigated samples in this work are MgO based pseudo spin valves. The CoFeB electrodes have an in-plane magnetic anisotropy. Generally, two methods are used to generate voltages across layer stacks by creating temperature gradients from top to bottom layer. The first one is by depositing an additional metallic layer on top of the stack and patterning a heater line directly above the MTJ [5, 10, 11]. Precise lithography methods allow the positioning of the heater line exactly on top of the MTJ, enabling full control of the location and direction of the temperature gradient. However, heating and cooling are rather slower, compared to optical excitations. Besides this, thermal insulation is necessary to avoid unwanted heating of other device regions. Furthermore, once a heater line is deposited, no further adjustments to the size and position can be undertaken. The second method is by using a laser [4, 12–14]. This approach appears more flexible, because the laser spot and thus the centre of the elevated temperature can be readjusted perpetually. However, for precise positioning of the MTJ with respect to the laser spot, motion steps with resolutions in the sub μm regime are required, considering MTJ and laser spot sizes below 10 μm in diameter. We found, that changes of the laser spot size and its position with respect to the MTJ lead to unwanted effects in the measurement signal. In general, positioning the centre of the laser spot not exactly in the centre of the MTJ, can lead to a smaller Seebeck voltage and even its reversal, resulting in a high TMS because of the vanishing voltage in the denominator in the calculation. That means, the high TMS ratios in those cases arise at the cost of signal because of a redistribution of the temperature gradients and thus small temperature differences between the electrodes.

---

[1] In other publications this term is also referred to as tunneling magneto thermopower TMTP or the magneto-Seebeck ratio $S_\mathrm{MS}$.

Here we present three series of measurements varying the properties of the heating laser beam. In the first series a variation of laser power is presented, which determines the optimum power for thermovoltage generation. In the second series of measurements, the influence of the laser spot size on the thermovoltage and the resulting TMS ratios is investigated. For increasing spot size, we find a drop of the Seebeck voltage, resulting in an increased TMS ratio. Consequently, the third series of measurements demonstrates the influence of systematically moving the laser beam in a square of the size 30μm × 30μm using steps of 1μm and recording thermovoltage maps locally. In this third series of measurements the origin of the signal change is identified. In both cases, when varying the spot size as well as when scanning the surface of the sample with the laser spot, the measured Seebeck voltages become a small amplitude and even change sign, when the MTJ is heated at the edges. Both effects are attributed to the resulting complex lateral temperature gradients direction change, which heats the bottom CoFeB electrode, keeping the top CoFeB electrode at a lower temperature. Furthermore, in-plane temperature gradients inside the MTJs play a significant role and should be investigated but have been difficult to access up to now. The aim of this study is to develop a technique to investigate these effects in more detail.

## 2. Methods
*2.1. Sample Fabrication*
The films are prepared under ultrahigh-vacuum conditions with a base pressure of $5 \times 10^{-10}$ mbar on MgO (100) substrates. This kind of substrate is chosen to avoid parasitic effects due to conductance from semiconducting substrates as described in reference [12]. The $Co_{20}Fe_{60}B_{20}$ electrodes and the Ta buffer layer are fabricated by magnetron sputtering. The MgO barrier was e-beam evaporated in a separate chamber without breaking the vacuum [15]. On top of the film stack, a Ru capping layer is deposited by e-beam evaporation to prevent oxidation of the underlying layers. The sample is *ex situ* annealed at a temperature of 450°C and an in-plane magnetic field of 300 mT for one hour. This procedure results in the crystallization of the amorphous CoFeB electrodes around the MgO layer and the diffusion of Boron into the Ta layers [16–18]. In the next step elliptical MTJs are patterned by electron beam lithography and subsequently ion-beam etched to a size of 6μm × 4μm with the long axis parallel to the direction of the magnetic field applied during the annealing procedure. Sputtered $Ta_2O_5$ (150 nm) is used as isolating material between individual MTJs. On top a 5-nm-thick Ta adhesion layer and a 70-nm-thick Au contact layer are deposited. The sample stack consists of Au 70 nm/ Ru 3 nm/ Ta 5 nm/ CoFeB 5.4 nm/ MgO 2.1 nm/ CoFeB 2.5 nm/ Ta 10 nm/ MgO substrate. A depiction of the layer stack and the structured junction is shown in figure 4c (left) and in figure 4c (right) the junction is shown embedded into the Au contactpads.

*2.2. Experimental setup and measurements*
In order to create a temperature gradient across the layer stack, we use a TOPTICA ibeam smart laser diode with a wavelength of 638 nm and $P \leq 150$ mW. The laser can be focused to a diameter $\geq 2$ μm full-width at half-maximum with a microscope objective (NIKON 20x, WD 20.5 mm). The position of the laser spot is controlled by a camera in a confocal microscope arrangement. The schematics of the setup are the same as in reference [12]. The thermovoltage is detected via a lock-in amplifier. A waveform generator (Agilent, 33500B series) modulates the laser diode with a square wave at a frequency of 77 Hz, which is used as modulation frequency for the lock-in amplifier. To measure magnetization dependent, the sample is situated in between two pole shoes of an electromagnet. Each thermovoltage versus external magnetic field curve is measured ten times and averaged. Linear stages with motorized actuators for the horizontal (x-direction) and vertical (y-direction) movement are implemented to position the laser beam on the sample surface. This function enables us to measure the thermovoltage versus magnetization direction by heating the sample at different positions over a defined area of 30μm × 30μm with $\leq 1$μm resolution, and additionally perform 2D scans. The laser spot diameter can be varied by adjusting the distance of the sample surface from the focusing objective (z-direction). The knife edge method is used to determine the beam width at the sample surface. In addition, the setup is equipped with a Keithley 2400 Sourcemeter to record TMR curves with a bias voltage of 10 mV to confirm the junction functionality.

*2.3. Tunnel-magnetoresistance measurements*
The electric resistance of a MTJ depends on the magnetic configuration of both magnetic electrodes. The effect describing the resistance change is the TMR. For parallel magnetization alignment of both electrodes, the resistance is low, for antiparallel alignment, the resistance increases [7, 8]. The TMR effect in the MTJ constitutes the necessary condition for the TMS effect, however, it is not the sufficient condition [4, 19]. Therefore, a high TMR effect does not guarantee a high TMS effect. The TMR ratio is measured before, in between and after the single TMS experiments. Figure 1a) shows, an example TMR curve recorded with a bias voltage of 10 mV for the junction investigated in this work. The resistance curve displays a sharp switching between 30 kOhm in the parallel and 104 kOhm in the antiparallel configuration of the ferromagnetic layers, this corresponds to a TMR ratio of 245 %. The curve confirms a clear separation of the magnetization states and the effect remains high. The TMR ratio is calculated according to:

$$TMR = \frac{R_{\text{ap}} - R_{\text{p}}}{R_{\text{p}}}.$$

*2.4. Tunnel magneto-Seebeck effect measurements*

For the TMS measurement the surface of the junction is illuminated by the laser beam thus a thermovoltage is generated. This voltage changes with the magnetization alignment of the magnetic CoFeB electrodes, similar to the resistance in the TMR measurements. A typical TMS effect measurement curve is depicted in figure 1b). The external magnetic field is swept beyond the coercive fields of the two CoFeB electrodes. For high external magnetic fields both electrodes are aligned parallel. Here the generated voltage differs from the voltage generated in the region where the magnetization alignment is antiparallel (indicated by the black arrows). The TMS ratio given on the right in figure 1b) is 55%. The TMS ratio was determined using:

$$TMS = \frac{V_{\text{ap}} - V_{\text{p}}}{\min(|V_{\text{ap}}|, |V_{\text{p}}|)}.$$

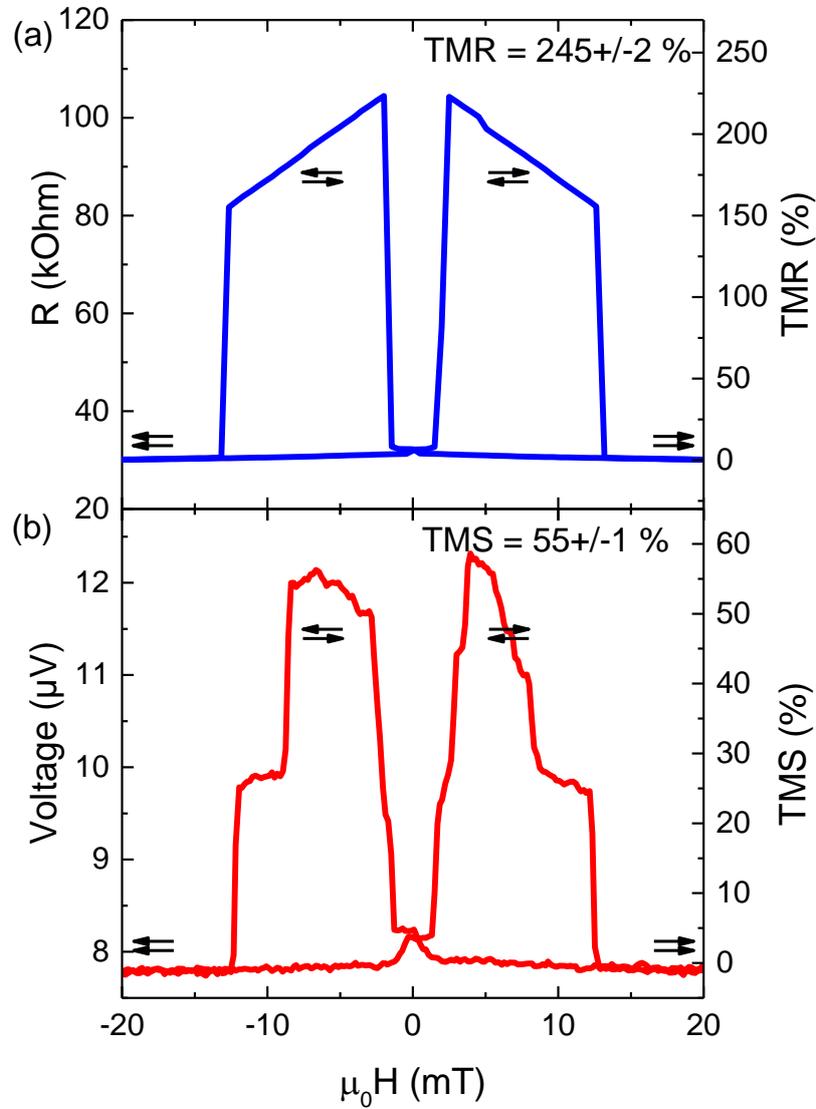

**Figure 1** Measurement curves of the (a) TMR effect, by measuring the resistance *R* of the MTJ and (b) the thermovoltage generated by the temperature gradient against the applied external field $\mu_0 H$. The arrows indicate parallel and antiparallel magnetization alignment of the electrodes, respectively. In case of the TMR measurement (a), applying a bias voltage of 10 mV, the resistance increase in the antiparallel alignment results in a TMR effect of 245%. In case of the TMS measurement (b), the thermovoltage also increases in the antiparallel alignment and results in a TMS effect of 55%.

## 3. Results and Discussion
*3.1. Power dependent TMS measurements*

Tuning the laser power, the temperature gradient is increased by depositing more energy in the top electrode and thus raising its temperature, while leaving the bottom electrode at the same base temperature guaranteed by the constant environmental settings of 22°C. The presented TMS measurements are performed with 10 mW, 25 mW and in 25 mW steps up to 150 mW laser power. The MTJ is heated in the centre to obtain the highest possible voltage that can be generated for each applied laser power. For the adjustment of the centre position, the laser power is tuned to 150 mW and the voltage signal is maximized by positioning the laser spot on the MTJ. The position of the laser spot remains unchanged throughout all measurements in this set. The measurements are performed setting the laser power randomly to avoid systematic errors in the following order: 100mW, 50mW, 10mW, 25mW, 75mW, 150mW, 125mW. For each setting the TMS is measured ten times and averaged. The extracted Seebeck voltages for the parallel and antiparallel magnetization alignment of both CoFeB electrodes versus the applied laser power are shown in figure 2 (left scale). On one hand the voltages for both orientations are increasing linearly with the laser power, the linear increase is indicated by the red and black lines, which are not fits to the data, but guides to the eye. In the parallel state values up to 7.6 µV are achieved and in the antiparallel state up to 12.3 µV at 150 mW laser power. On the other hand, the calculated TMS values depicted in figure 2 (right scale) as a function of laser power, remain at a constant rate between 65% and 55%, as indicated by the blue line as a guide to the eye. This indicates, that effects heating the MTJs surroundings at high laser powers in the given range are not present. High laser power focused to one spot and centreed on top of the junction creates a temperature gradient with a high temperature difference between the top and bottom electrodes. Because the power dependent measurements show the highest voltage signal for 150 mW, all further measurements are performed by keeping the laser power at this value.

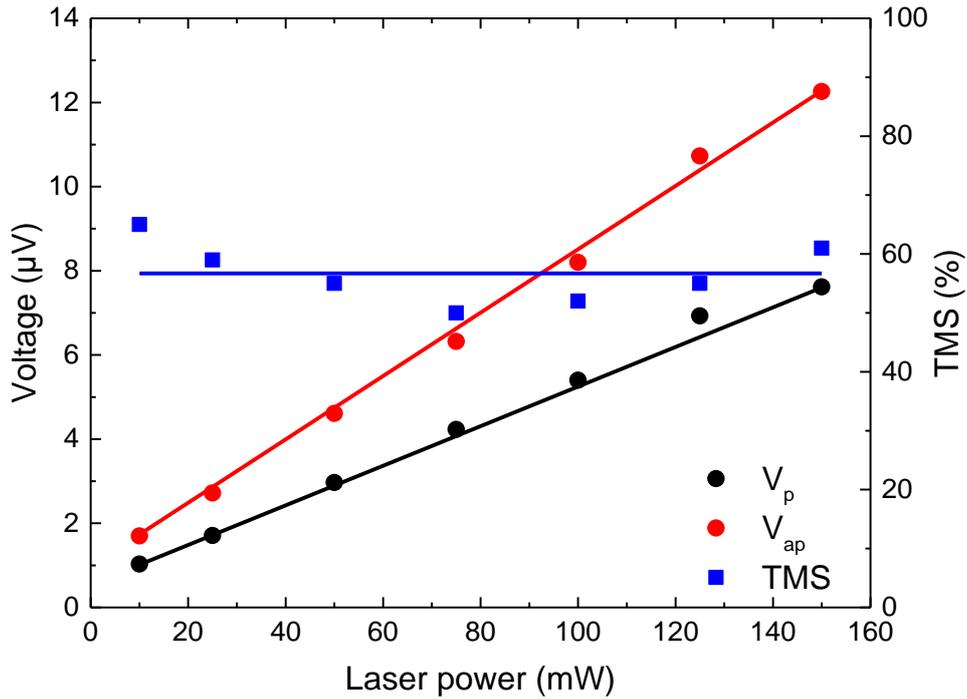

**Figure 2** The extracted voltages for the parallel magnetization alignment $V_p$ (black dots) and the antiparallel magnetization alignment (red dots) of the CoFeB electrodes (left scale). The resulting TMS ratio is calculated from the voltages (blue squares, right scale). Both quantities are plotted as functions of the laser power applied for heating. All depicted lines are guides to the eye.

*3.2. Laser spot size dependent TMS measurements*

For a further investigation the heat gradient is varied, by increasing the laser spot size $d$ on the tunnel junction element of 6μm × 4μm. The first step of the procedure is to place the sample into the focal point of the objective. Here, the spot size is 2.7μm. Moving the sample away from the objective in z-direction, the spot size increases to 6.7μm. The spot size for each z-position of the sample is determined using the knife edge method, by moving the sample horizontally in x-direction and recording the signal change of the reflected light at the edge of the Au contact pad and the $Ta_2O_5$ insulator. The spot size is calculated from the recorded data according to reference [20], the laser beam profile is circular. In order to heat the MTJ directly on top and to create a temperature difference preferably only within the MTJ, a spot size smaller than the actual MTJ is chosen. Increasing the spot size, also a temperature difference is created in the surroundings of the MTJ. The extracted thermovoltages for the parallel magnetization alignment $V_p$, the antiparallel magnetization alignment $V_{ap}$ and the voltage difference $|V_{ap} - V_p|$ are plotted in figure 3 (left scale). The calculated TMS is plotted in figure 3 (right scale). The voltage decreases for both magnetization alignments proportional to $\frac{1}{d^2}$, since this is also the factor by which the laser fluence is reduced when the laser spot diameter increases. Additionally, for spot sizes, that exceed the MTJ dimensions, $V_p$ becomes even negative. There are two effects leading to small thermovoltages. The first is, that with increasing the laser spot size, the energy density deposited at the sample surface decreases. This leads to a lower temperature rise in the top CoFeB electrode in respect to the bottom electrode, because the energy is distributed over a larger in-plane area, and thus generating a smaller thermovoltage. The second is in the case of the sample design used here. We assume that the $Ta_2O_5$ layer is transparent for the incoming photons, which are transmitted through the Au layer to the bottom Ta layer [21]. However, the intensity transmitted through the Au and top Ta layer is below 1%, considering the thickness of 70nm. This would transport less than 1mW of the laser power to the bottom layers, which is not enough to generate a thermovoltage across the MTJ. Besides this, theoretical simulations of the Seebeck coefficients do not show a sign change in the temperature regions above 300K, even in view of variations in the Co and Fe compositions of the electrodes [19]. However, looking at the patterned MTJ structures at the edges, like it was investigated in reference [22], the etching process does not homogenously remove the material, shaping straight pillars, but instead, the bottom electrode becomes broad. At the interface of the pillar and the $Ta_2O_5$ insulator, the energy can be passed on to the bottom CoFeB electrode, raising its temperature and, thus, reversing the temperature gradient inside the MTJ. The calculated TMS ratios increase from around 50% to above 1300% to the point, where $V_p$ changes sign. The drastic increase of the TMS ratio in this region is caused by the rapid decrease of $V_p$, with increasing laser spot size, that is stronger than the decrease of $|V_{ap} - V_p|$. Also, too small laser spot sizes will lead to unwanted in-plane temperature gradients within the MTJ. In figure 3, the green area represents the spot size range that is best adapted to the junction size. It has to be noted that for the present sample layout of MTJs, pattern structure and oxide used for electric isolation, those in-plane effects are most prominent. Although a sign change of the thermovoltage is not always observed, it shows that homogenous heating of the whole MTJ is crucial to obtain reliable values of the thermovoltage and the TMS ratio.

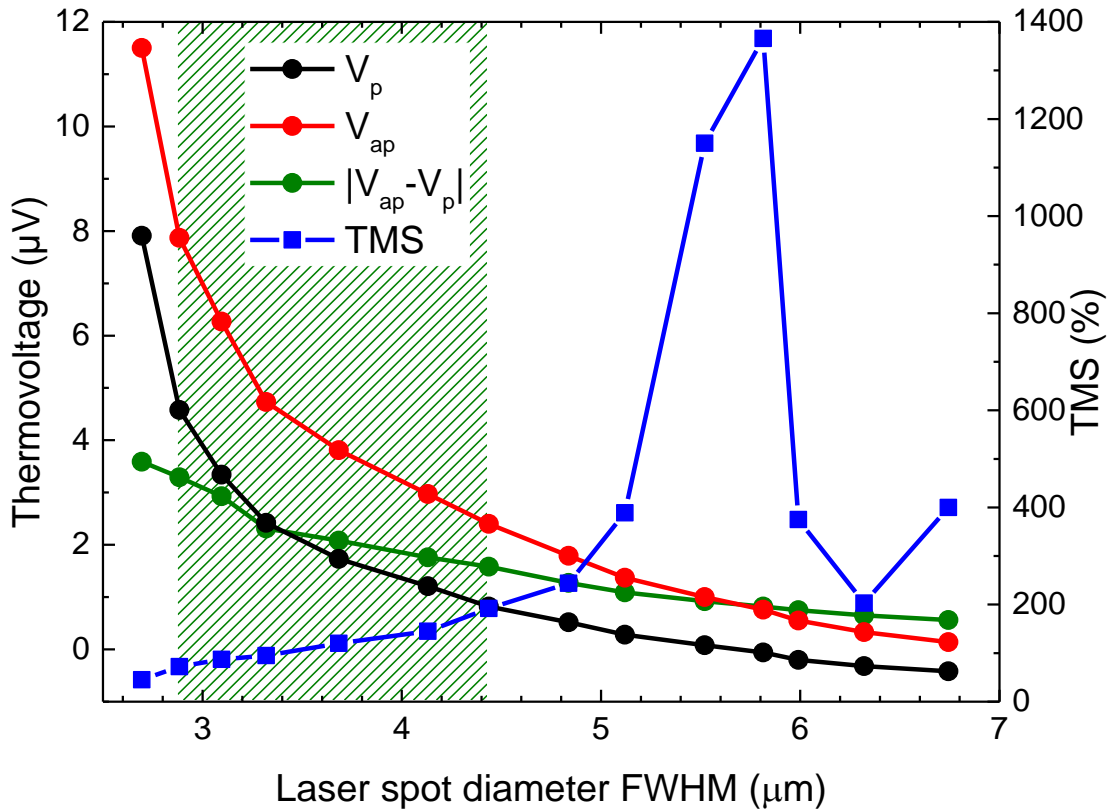

**Figure 3** Dependence of measured thermovoltages and TMS ratios on the laser spot diameter for an elliptical shaped MTJ (6µm x 4µm) with a laser power of 150mW. The left scale shows the extracted voltages for the parallel magnetization alignment $V_p$(black dots), the antiparallel magnetization alignment $V_{ap}$(red dots) and the difference $|V_{ap} - V_p|$. The right scale shows the TMS ratio calculated from the measured voltages. All data is plotted as a function of the laser spot size variation. The green shaded area displays the range with most homogenous heating, resulting in a well-defined out-of-plane temperature gradient. Here the diameter is adapted to the size of the given MTJ.

*3.3. Two-dimensional scan of the magnetic tunnel junction*

In order to deepen the understanding of the effects observed and confirm the conclusions made, position dependent laser spot measurements are performed. For this set of measurements, the laser spot size is set to 2.6µm. This ensured a small heating area and high Seebeck voltages when heating the junction in the centre. To define the scanning area, the laser spot was positioned in the centre of the MTJ. Then the MTJ was moved 15µm in both, the x- and the y-direction. From that point, the MTJ was moved in 1µm steps a distance of 30µm in both directions towards the MTJ, recording the magnetization dependent Seebeck voltages like shown in figure 1b at each point. The extracted voltages for the parallel (Figure 5a) and the antiparallel (Figure 5b) magnetization alignment are plotted in three-dimensional graphs. Two dimensions are used to depict the scanned area (x- and y-direction). The third dimension (z-scale) displays the recorded values. Additionally the same data is depicted zoomed in in a false color plot (Figure 5c, 5d). The figures demonstrate, that the generated voltage decreases exponentially, as the laser spot is moved away from the centre of the MTJ. In close vicinity of the MTJ, when the laser beam is not striking it anymore, the measured voltage becomes negative, before the signal goes to 0, at larger distance about 15µm away from the junction centre. The false color plots show a further analysis, from which the position and the shape of the MTJ can be reconstructed, by fitting a two dimensional Gaussian function to the data. The fits reveal a rotation of the MTJs long axis of $\sim 10°$ from the vertical position, which is reasonable, considering inaccuracies when installing the sample in the setup. Further, the FWHM dimensions of the fits have dimensions of 3.9 µm and 4.6 µm for the short axis and 4.7µm and 6.1µm for the long axis. Those are the dimensions of the junction. This confirms, that the best signal can only be achieved, when heating the junction in the centre. Altogether, the voltages, extracted for the parallel and antiparallel magnetization alignment, indicate, which parameters need to be satisfied in order to achieve high signals. First, the voltage increases linearly with the laser intensity applied. Second, the laser power needs to be focused to the highest energy density. Third, the MTJ needs to be heated by the laser beam in the centre. The calculated TMS ratios from the extracted voltages for the parallel and antiparallel magnetization alignment are plotted in figure 6. The results are similar to those discussed in section 3.2. In both cases, when heating the vicinity of the MTJ, the measured voltages decrease and even change their sign, the TMS ratio increases and reaches values up to 6000%.

The cross-section along the horizontal 0 line from figure 5c) is presented in figure 4a), the thermovoltages $V_p$, moving the laser beam across the centre of the MTJ along the short axis. The voltage values are given by the red dots connected to the x-axis by drop lines. The grey area at the bottom of the graph represents the $Ta_2O_5$ with the MTJ layer stack in between. For a clarification of the sample structure a sketch of the MTJ layer stack enlarged, and embedded into the Au and Ta contact pads is depicted in figure 4c). The Gaussian curve indicating the heat distribution across the layer stack and its in-plane expansion is drawn into the plot to illustrate the direction of the incoming laser beam onto the sample and the resulting perpendicular temperature gradient from the top to the bottom layer. The data clearly states, that the largest temperature difference between the top and the bottom layer in the junction is achieved, when the laser spot is positioned in the centre of the MTJ. Accordingly, the highest thermovoltages are generated, positioning the laser beam within 2.5µm distance from the centre of the MTJ. At distances more than 8µm away from the centre of the MTJ, the temperature remains largely unaffected. Considering the optical properties [21, 23, 24] of $Ta_2O_5$, our first interpretation of the sign change is that the heat is transported through the $Ta_2O_5$ layer to the Ta bottom contact layer, which is distributed throughout the whole sample and as a result only the temperature of the bottom CoFeB layer increases. However, as stated in section 3.2, the laser power transmitted through the Au layer does not exceed 1%. Taking into account the low thermal conductivity and high heat capacity of $Ta_2O_5$ [25, 26], the heat is not conducted to the bottom Ta layer, which could pass it on to the bottom CoFeB electrode. This confirms the interface between the pillar stack and the $Ta_2O_5$ as the only channel to transport the heat to the bottom CoFeB electrode, which because of the irregularities remaining form the patterning process leads to a reversal of the temperature gradient. Because the in-plane distance is one order of magnitude larger than the layer stack thickness, the temperature difference is low. This leads to only a very small negative voltage. When positioning the laser spot 2μm away from the MTJs edge, the voltage is small, but the calculated TMS ratio reaches values higher than 1000 %. In this region, the layer stack is heated laterally, resulting in an elevated temperature, but only to a small temperature gradient and thus a small voltage is generated. To visualize this, figure 4b) shows the thermovoltage measurement curves numbered 1-8, from which the voltages in figure 4a) were extracted. Those curves show clearly, how the difference between the voltage in the parallel and antiparallel magnetization alignment decreases, when the laser is moved over the sample surface. In addition, it is also clearly visible, how the signal altogether decreases. These small signals give rise to the disproportionate increase in the calculated TMS ratios.

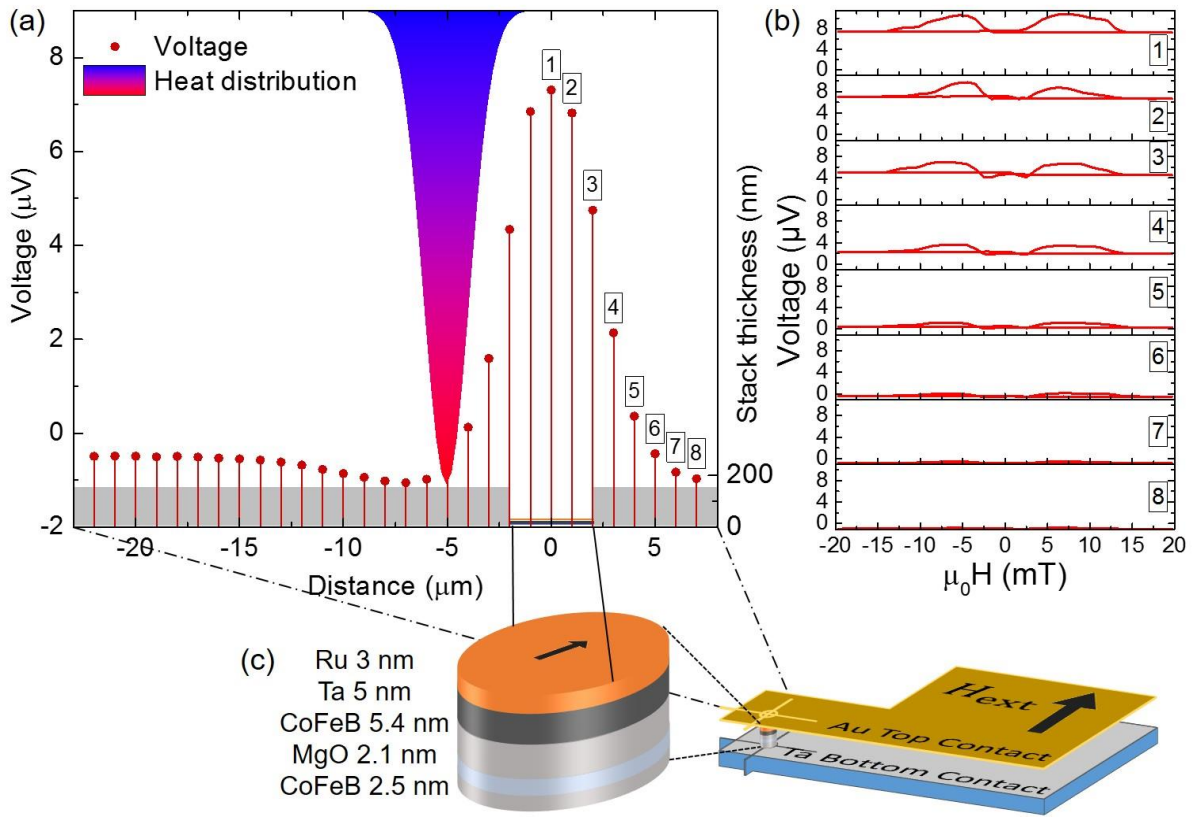

**Figure 4** (a) Cross-section of the voltages recorded along the centre of the MTJs short axis, taken from the horizontal 0-line in figure 5a) (left scale). The grey shaded area at the bottom represents a cross-section of the sample (150 nm $Ta_2O_5$) with the MTJ layer stack in between. The thickness is indicated on the right scale. The blue-red Gaussian function indicates the vertical heat gradient. The voltages numbered 1-8 are extracted from measurement curves in (b). (c) The structure of the pure MTJ is sketched on the left. The MTJ is embedded in the Au and Ta contacts. The blue layer underneath represents the substrate. The arrows show the direction of the external field.

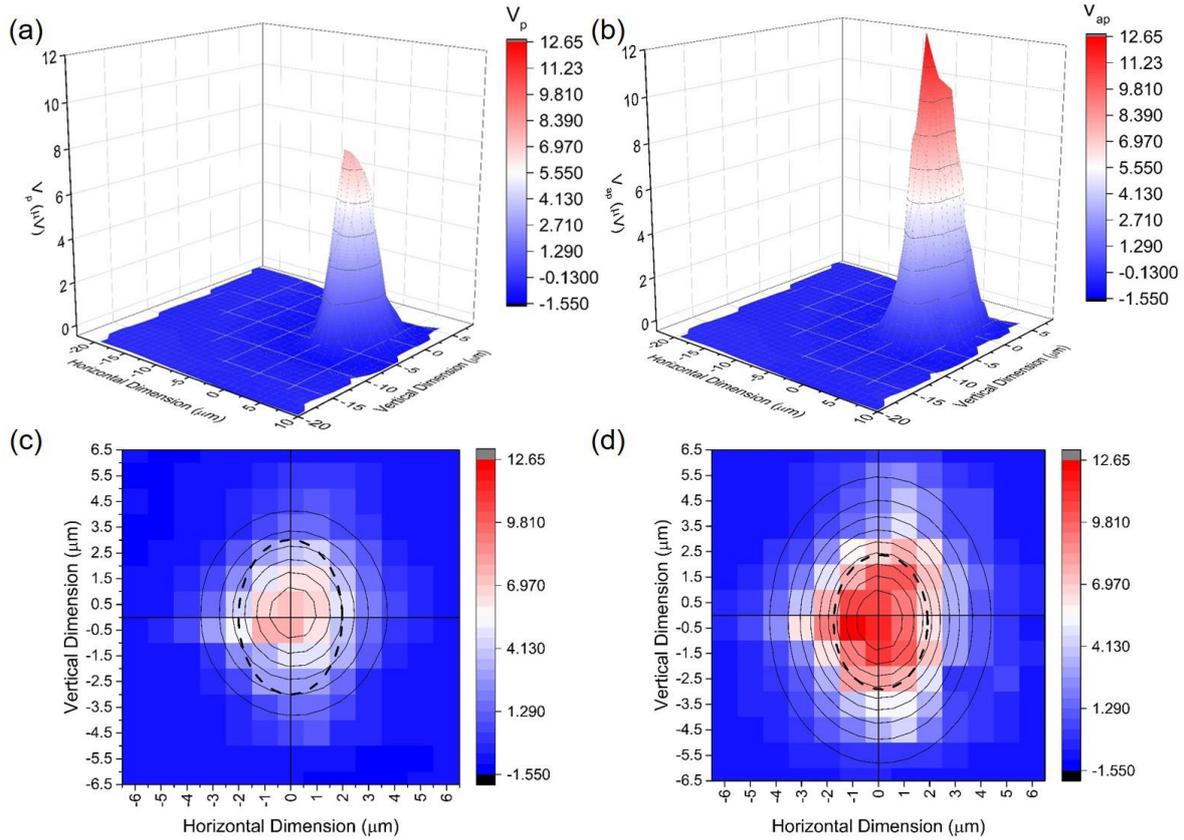

**Figure 5** The extracted thermovoltages for the (a) parallel $V_p$ and (b) antiparallel $V_{ap}$ magnetization alignment as a function of the laser spot position (0 position is the centre of the MTJ). The false color plots in (c) and (d) show the sections of $V_p$ and $V_{ap}$ respectively centreed on the voltage peaks. In addition, both graphs contain two dimensional Gaussian function fits to the data, outlining the decrease of the signal from the centre to the edge. The dashed centreed ellipse has the dimensions of the MTJ (6μm × 4μm).

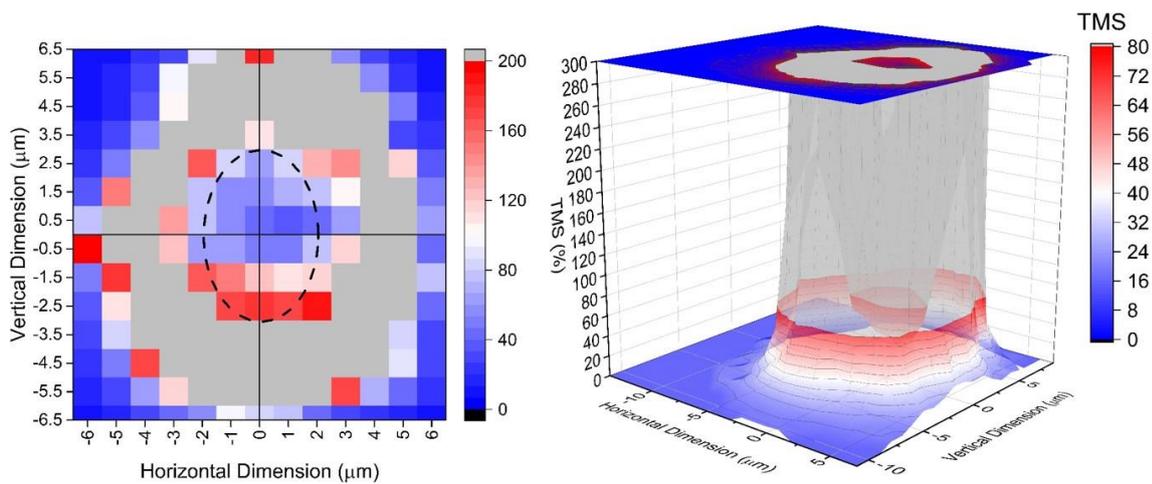

**Figure 6** The TMS ratio calculated from the data depicted in the two dimensional scan (0 Position is the centre of the MTJ). On the left, the false color plot shows the region centreed on the MTJ, marked by the dashed ellipse. The greyed area depicts TMS ratios larger than 200%. The three dimensional

surface plot illustrates the centre, where the MTJ is located with moderate TMS ratios around 50% - 70%.

## 4. Conclusion

In this work, we have systematically varied parameters of the laser beam to demonstrate effects of the lateral profile of the created temperature gradient in an MTJ layer stack and its influence on the generated thermovoltage and the corresponding TMS ratio. For the type of junction and oxide used for electric isolation we find that the best results are obtained, when the laser beam is situated directly above the centre of the MTJ and its diameter are smaller and equal to the junction size. That concludes, that for those laser spot sizes edge heating effects are minimized, reducing the in-plane created temperature gradients, leading to most homogeneous heating in the direction across the layer stack. The apparently high TMS ratios arise from lateral temperature gradients, which lead to elevated temperatures, but only small temperature gradients. These are mirrored in the small measurement signals leading to large TMS values. The thermovoltage maps show that a controlled temperature gradient has to be applied to avoid misinterpretations of the TMS effect. These effects of spot size and spot positioning are especially important if a sign change is observed in the lateral voltage maps. Furthermore, this shows an additional aspect, when comparing the data to theoretical simulations of the Seebeck coefficients. Here not only the temperature difference, that so far is only accessible through heat transfer simulations, but also the sensitivity of the thermovoltage signal in respect to the created temperature profiles can cause fluctuations in the calculated Seebeck coefficients. However, it also shows that the TMS in nanoscale spintronic devices allows much more variation of voltage and thermal landscapes in spincaloritronic devices on the micrometer scale.

Acknowledgements. This work was funded by the German Research Foundation (DFG) through the priority program `SpinCaT` (SPP 1538).